\documentclass[aps,twocolumn]{revtex4}
\usepackage{epsfig}
 \usepackage{pdfpages}
\usepackage{youngtab}
\usepackage{graphicx}
\usepackage{amsmath}
\usepackage{booktabs}
\usepackage{color}

\begin{document}
\title{Compact hidden charmed pentaquark states and QCD isomers}
\author{Cheng-Rong Deng$^{a,b}{\footnote{crdeng@swu.edu.cn}}$}

\affiliation{$^a$School of Physical Science and Technology, Southwest University, Chongqing 400715, China}
\affiliation{$^b$School of Physics and Center of High Energy Physics, Peking University, Beijing 100871,China}

\begin{abstract}
We make an exhaustive investigation on the pentaquark states $qqqc\bar{c}$ ($q=u, d$ and $s$) and discuss the
effect of color structures in a multiquark color flux-tube model. We exhibit a novel picture of the structure
and properties of the states $P_c$ and $P_{cs}$ observed by the LHCb Collaboration. We can describe the states
as the compact pentaquark states in the model. The spin-parity of the group of $P_c(4312)^+$ and $P_c(4337)^+$
is $\frac{1}{2}^-$ while that of the group of $P_c(4380)^+$, $P_c(4440)^+$ and $P_c(4457)^+$ is $\frac{3}{2}^-$.
Their structures are pentagon, diquark, pentagon, diquark, and octet, respectively. The members in each
group can be analogically called QCD isomers because of their the same spin-parity and quark content but different
color structures. The singlet $P_{cs}(4459)^0$ has pentagon structure and spin-parity of $\frac{1}{2}^-$.
In addition, we also predict the $P_{cs}$, $P_{c ss}$ and $P_{csss}$ families in the model. The five-body
confinement potential based on the color flux-tube picture, which is a collective degree of freedom and induces QCD
isomer phenomenon, plays an important role in the formation of the compact states.

\end{abstract}

\maketitle

\section{Introduction}

Conventional baryons are composed of three valence quarks in the constituent quark models.
Exploring exotic baryons consisting of four valence quarks and one valence antiquark, called
pentaquark states, has been one of the most significant research topics in the hadron physics
since the birth of the quark model~\cite{history}. The existence of fully-light pentaquark
states is apt to be negative so far~\cite{light}. In the charm sector, there were many
predictions on the hidden charmed pentaquark states~\cite{wjju,wlwang,jjwu1,cwxiao,karliner}.
Recently, the LHCb Collaboration reported the hidden charmed pentaquark states $P_c(4380)^+$,
$P_c(4312)^+$, $P_c(4440)^+$, $P_c(4457)^+$, $P_{cs}(4459)^0$, and $P_c(4337)^+$ in the $J/\psi p$
or $J/\psi \Lambda$ invariant mass spectrum~\cite{pc43804450,pc4312,pcs4459,pc4337}. Their masses,
widths and minimal valence quark contents are presented in Table~\ref{pcs}. However, the reliable
information about their spin-parity has been unavailable until now.
\begin{table}
\caption{The states $P_c$ and $P_{cs}$.}\label{pcs}
\begin{tabular}{ccccccccccc}
\toprule[0.8pt] \noalign{\smallskip}
       State                      &             Mass (MeV)              &         Width (MeV)             &   Content     \\
\toprule[0.8pt] \noalign{\smallskip}
$P_c(4380)^+$~\cite{pc43804450}   &            $4380\pm8\pm29$          &       $215\pm18\pm86$           & $uudc\bar{c}$ \\
\noalign{\smallskip}
$P_c(4312)^+$~\cite{pc4312}       & ~~$4311.9\pm0.7^{~+6.8}_{~-0.6}$~~  &  $9.8\pm2.7^{~+3.7}_{~-4.5}$    & $uudc\bar{c}$ \\
\noalign{\smallskip}
$P_c(4440)^+$~\cite{pc4312}       &    $4440.3\pm1.3^{~+4.1}_{~-4.7}$   &  $20.6\pm4.9^{~+8.7}_{~-10.1}$  & $uudc\bar{c}$ \\
\noalign{\smallskip}
$P_c(4457)^+$~\cite{pc4312}       &    $4457.3\pm0.6^{~+4.1}_{~-1.7}$   &  $6.4\pm2.0^{~+5.7}_{~-1.9}$    & $uudc\bar{c}$ \\
\noalign{\smallskip}
$P_{cs}(4459)^0$~\cite{pcs4459}   &    $4458.8\pm2.9^{~+4.7}_{~-1.1}$   &  $17.3\pm6.5^{~+8.0}_{~-5.7}$   & $udsc\bar{c}$ \\
\noalign{\smallskip}
$P_c(4337)^+$~\cite{pc4337}       &       $4337^{~+7~+2}_{~-4~-2}$      &  $29^{~+26~+14}_{~-12~-14}$     & $uudc\bar{c}$ \\
\noalign{\smallskip}
\toprule[0.8pt]
\end{tabular}
\end{table}

Systematical study on their nature and structure can improve our understanding of the non-perturbative
behaviors of the strong interaction. Therefore, a lot of theoretical explanations have been devoted
to their properties, such as hadron molecular states~\cite{moleculedu,moleculewang,4312moleculehqss,moleculehxchen,moleculerchen},
compact pentaquark states~\cite{compactsantopinto,4380cftm,compactrlzhu,compactlebed,compactali,4312stancu},
kinematical effects~\cite{kineticguo,kineticliu}, and hadrocharmonium~\cite{hadrocharmonium1,hadrocharmonium2},
virtual states~\cite{unbound}, and double triangle cusps~\cite{cusp}, within differen theoretical frameworks.
The latest reviews can be found in Refs.~\cite{reviews}, in which the molecular state is overwhelming
because of the proximity of their masses to the baryon-meson thresholds. Even so, there were no conclusive
consensus on their properties, especially for the states $P_c(4440)^+$ and $P_c(4457)^+$ because
of their ambiguous spin~\cite{noconcensus}.

The color structures of mesons and baryons are unique while the multi-quark states have abundant
color structures~\cite{lattice1,lattice2}. The effect of various color structures, which is absent
in the mesons and baryons, may raise in the multiquark states. The states $P_c$ and $P_{cs}$
provide a good platform to explore the effect. In the previous work~\cite{4380cftm}, we studied
the state $P_c(4380)^+$ and proposed a novel color flux-tube structure, a pentagon state, for the
pentaquark states in the multiquark color flux-tube model. In the present work, we prepare to make
a systematical investigation on the hidden-charm pentaquark states in the model. We anticipate to
exhibit new insights into the properties and structures of the $P_c$ and $P_{cs}$ states from the
perspective of the phenomenological model. We also hope that this work can improve the understanding
of the mechanism of the low-energy strong interactions.

This paper is organized as follows. After the introduction, Sec. II gives the descriptions of the
model. Sec. III presents the wave functions of the hidden charm pentaquark states. Sec. IV shows
the numerical results and discussions. The last section lists a brief summary.

\section{Multiquark color flux-tube model (MCFTM)}

Lattice QCD investigations on mesons and baryons revealed their internal color structures~\cite{lattice1},
see FIG. 1. The quark and antiquark in mesons are linked with a three-dimensional color flux tube.
Three quarks in baryons are connected by a Y-shape flux-tube, in which $\mathbf{y}_0$ denotes
a junction where three color flux tubes meet.

The hidden charmed pentaquark states have four possible color flux-tube structures~\cite{4380cftm},
(1) meson-baryon molecular state (molecule), (2) diquark-diquark-antiquark state (diquark),
(3) color octet state (octet), and (4) pentagonal state (pentagon), which are shown in FIG. 2.
The corresponding positions of quarks and antiquark are denoted as $\mathbf{r}_1$, $\mathbf{r}_2$,
$\mathbf{r}_3$, $\mathbf{r}_4$, and $\mathbf{r}_5$, $\mathbf{y}_i$ represents the $i$-th Y-shape
junction. In some extent, color flux-tube is similar to chemical bond in QED. The QED isomers 
have same atom constituents but different chemical bond structures. Analogously, we can call such
different structures QCD isomers.
\begin{figure} [h]
\begin{center}
\resizebox{0.35\textwidth}{!}{\includegraphics{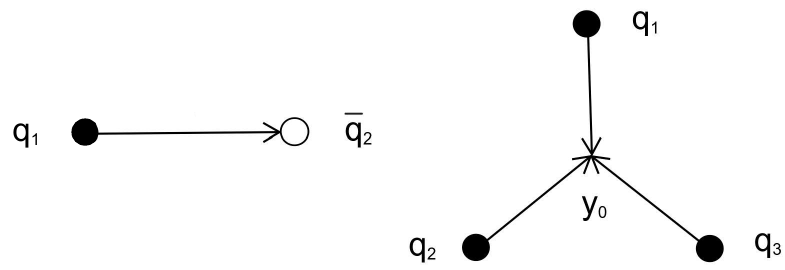}}
\caption{Meson and baryon states.}
\label{2q-3q}
%\end{figure}
%\begin{figure} [h]
\smallskip\smallskip
\resizebox{1.15\textwidth}{!}{\includegraphics{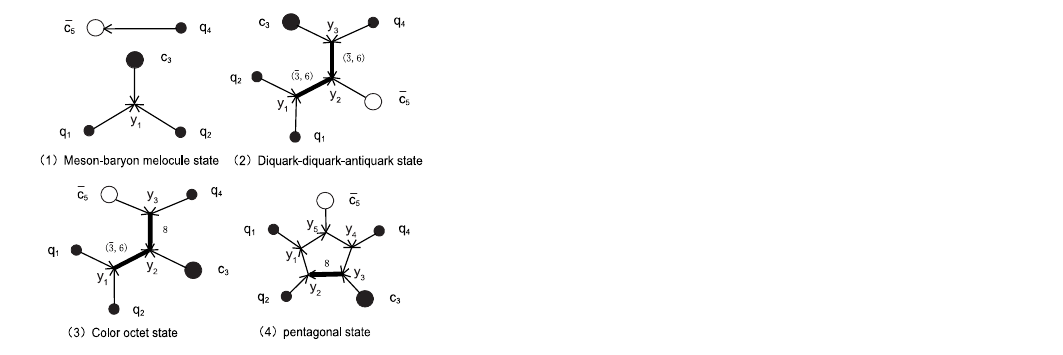}}
\caption{Hidden charm pentaquark states.}
\label{flux-tube}
\end{center}
\end{figure}

A thin line only stands for a $\mathbf{3}$- or $\bar{\mathbf{3}}$- dimension color flux-tube 
while a thick line represents a $\mathbf{3}$-, $\bar{\mathbf{3}}$-, $\mathbf{6}$-, 
$\bar{\mathbf{6}}$- or $\mathbf{8}$-dimension color flux-tube. The arrow represents the 
color coupling direction. Two color flux tubes meet at a Y-shape junction along with the 
direction of the arrows, where the coupling of two colors carried by the color flux tubes 
into another color carried by the third color flux tube starting from the Y-shape junction, 
such as the three color flux tubes $q_1\mathbf{y}_1$, $q_2\mathbf{y}_1$ and 
$\mathbf{y}_1\mathbf{y}_2$ in FIG. 2 (2) and (3), $\mathbf{3}\otimes\mathbf{3}
=\bar{\mathbf{3}}\oplus\mathbf{6}$. Three color flux tubes meet a Y-shape junction along 
with the direction of the arrow, such as $\mathbf{y}_1\mathbf{y}_2$, $\mathbf{y}_3\mathbf{y}_2$ 
and $\bar{c}_5\mathbf{y}_2$ in the diquark structure, where three colors can couple into 
a color singlet. In this way, the connection between the color flux-tube structure and 
the construction of the color wave function can be established clearly.

The construction of the color wave functions, no matter baryons, the pentaquark states 
with diquark or octect configurations, is based on the degrees of quark freedom. In 
another word, its starting point is always the color coupling of quark-quark or 
quark-antiquark in one Y-shape color flux-tube, such as $q_1$-$q_2$ in FIG. 2 (2) or
$q_4$-$\bar{c}_5$ in FIG. 2 (3). However, none of quark-quark or quark-antiquark 
is in one Y-shape color flux-tube in FIG. 2 (4). Any two quarks are connected by 
two or more Y-shape color flux-tubes. Therefore, how to establish its color wave 
functions is an open question in the quark level. Even so, this ringlike structure 
does not violate QCD and it can form an overall color singlet. Richard also explored
similar ringlike structure of hexaquark states in the string model~\cite{richard}. 
In the present work, we first apply the wave function of the diquark structure to 
estimate the energy of the pentagon structure approximately.

The MCFTM has been established on the basis of the traditional quark models and 
lattice QCD color flux-tube picture~\cite{4380cftm,cftm}. Comparing with the 
traditional constituent quark models, the MCFTM merely modify the sum of two-body
confinement potential in the traditional models to a multi-body quadratic one. 
Relative to the lattice QCD, we replace the linear potential with the quadratic 
one. For the ground hadron states, their sizes are generally less than or around 
1 fm, in which the difference between the quadratic potential and the linear one 
is not obvious. The difference can be further diluted by the adjustable stiffnesses
of color flux-tube. The replacement is therefore reasonable in the ground states. 
Note that the replacement in the excited states needs to be addressed with great 
caution because they are spatially more extended ($>$1 fm). In addition, the 
quadratic confinement potential can greatly simplify the numerical calculation 
in the dynamical investigation on the multiquark states.

In the MCFTM, the two-body quadratic confinement potential for mesons can be written as
\begin{eqnarray}
V_{min}^{con}(2)=k(\mathbf{r}_1-\mathbf{r}_2)^2,
\end{eqnarray}
where $k$ is the stiffnesses of a three-dimension color flux-tube. The three-body quadratic
confinement potential for baryons can be written as
\begin{eqnarray}
V^{con}(3)=k\sum_{i=1}^3(\mathbf{r}_i-\mathbf{y}_0)^2
\end{eqnarray}
We can determine the junction $\mathbf{y}_0$ of the Y-shape structure by taking the variation 
on the three-body quadratic confinement potential, 
\begin{equation}
\mathbf{y}_0=\frac{\mathbf{r}_1+\mathbf{r}_2+\mathbf{r}_3}{3}.
\end{equation}
Then we can arrive at the minimum of the confinement potential for baryons,
\begin{equation}
V_{min}^{con}(3)=
k\left(\left(\frac{\mathbf{r}_1-\mathbf{r}_2}{\sqrt{2}}\right)^2
+\left(\frac{2\mathbf{r}_3-\mathbf{r}_1-\mathbf{r}_2}{\sqrt{6}}\right)^2\right).
\end{equation}

According to the color flux-tube structures of the hidden charmed pentaquark states in
FIG. 2, the confinement potential of the $i$-th color structure $V_i^{con}(5)$ reads
\begin{widetext}
\begin{eqnarray}
V^{con}_1(5)&=& k\sum_{i=1}^3(\mathbf{r}_i-\mathbf{y}_1)^2+k(\mathbf{r}_4-\mathbf{r}_{5})^2,\\
V^{con}_2(5)&=& k\sum_{i=1}^2(\mathbf{r}_i-\mathbf{y}_1)^2+k\sum_{i=3}^4(\mathbf{r}_i-\mathbf{y}_3)^2 +k\sum_{i=1,3}(\mathbf{y}_i-\mathbf{y}_2)^2+k(\mathbf{r}_{5}-\mathbf{y}_{2})^2,\\
V^{con}_3(5)&=&k\sum_{i=1}^2(\mathbf{r}_i-\mathbf{y}_1)^2+k\sum_{i=4}^5(\mathbf{r}_i-\mathbf{y}_3)^2
+k\left((\mathbf{y}_{1}-\mathbf{y}_{2})^2+\kappa_8(\mathbf{y}_{2}-\mathbf{y}_{3})^2+(\mathbf{r}_{5}-\mathbf{y}_{2})^2\right),\\
\noalign{\smallskip}
V^{con}_4(5)&=&k\sum_{i=2,5}(\mathbf{y}_i-\mathbf{y}_{1})^2+k\sum_{i=3,5}(\mathbf{y}_i-\mathbf{y}_{4})^2+k\sum_{i=1}^5(\mathbf{r}_i-\mathbf{y}_{i})^2+\kappa_8 k(\mathbf{y}_2-\mathbf{y}_{3})^2
\end{eqnarray}
\end{widetext}
where $\kappa_{d}k$ is the stiffness of the $d$-dimension color flux-tube, $\kappa_d=\frac{C_{d}}{C_3}$~\cite{kappa}.
$C_{d}$ is the eigenvalue of the Casimir operator associated with the $SU(3)$ color representation $d$ at either
end of the color flux-tube.

We can obtain the junctions $\mathbf{y}_i$ by taking the variation on each five-body quadratic confinement potential.
Then, we achieve the eigenvectors $\mathbf{\xi}_i$, $\mathbf{\chi}_i$, $\mathbf{\zeta}_i$ and $\mathbf{\eta}_i$ and
their corresponding eigenvalues by diagonalizing the confinement potential matrixes. The eigenvectors are
in fact the normal modes of the five-body quadratic confinement potentials, which read
\begin{widetext}
\begin{center}
\begin{eqnarray}
\left (
\begin{array}{cccccc}
 \mathbf{\xi}_1   \\
 \noalign{\smallskip}
 \mathbf{\xi}_2    \\
 \noalign{\smallskip}
 \mathbf{\xi}_3      \\
 \noalign{\smallskip}
 \mathbf{\xi}_4       \\
 \noalign{\smallskip}
 \mathbf{\xi}_5    \\
\end{array}
\right )=\left (
\begin{array}{cccccc}
\frac{1}{\sqrt{2}}  &\frac{-1}{\sqrt{2}}   &        0             &             0         &         0             \\
\noalign{\smallskip}
        0           &           0            &\frac{-1}{\sqrt{2}} &   \frac{1}{\sqrt{2}}  &         0             \\
\noalign{\smallskip}
\frac{1}{\sqrt{6}} & \frac{1}{\sqrt{6}} &\frac{-2}{\sqrt{6}} & 0 & 0       \\
\noalign{\smallskip}
\frac{\sqrt{2}}{\sqrt{15}} &\frac{\sqrt{2}}{\sqrt{15}}  & \frac{\sqrt{2}}{\sqrt{15}}  & \frac{-\sqrt{3}}{\sqrt{10}} & \frac{-\sqrt{3}}{\sqrt{10}}  \\
\noalign{\smallskip}
\frac{1}{\sqrt{5}}  &    \frac{1}{\sqrt{5}}  &  \frac{1}{\sqrt{5}}  &   \frac{1}{\sqrt{5}}  &  \frac{1}{\sqrt{5}}
\end{array}
\right )\left (
\begin{array}{cccccc}
\mathbf{r}_1   \\
\noalign{\smallskip}
\mathbf{r}_2    \\
\noalign{\smallskip}
\mathbf{r}_3      \\
\noalign{\smallskip}
\mathbf{r}_4       \\
\noalign{\smallskip}
\mathbf{r}_5    \\
\end{array}
\right ),~~
\left (
\begin{array}{cccccc}
\mathbf{\chi}_1   \\
\noalign{\smallskip}
\mathbf{\chi}_2    \\
\noalign{\smallskip}
\mathbf{\chi}_3      \\
\noalign{\smallskip}
\mathbf{\chi}_4       \\
\noalign{\smallskip}
\mathbf{\chi}_5
\end{array}
\right )=\left (
\begin{array}{cccccc}
\frac{1}{\sqrt{2}}  &  \frac{-1}{\sqrt{2}}   &        0             &             0         &         0             \\
\noalign{\smallskip}
        0           &           0            &  \frac{-1}{\sqrt{2}} &   \frac{1}{\sqrt{2}}  &         0             \\
\noalign{\smallskip}
\frac{1}{\sqrt{4}}  &   \frac{1}{\sqrt{4}}   &  \frac{-1}{\sqrt{4}} &   \frac{-1}{\sqrt{4}} &         0             \\
\noalign{\smallskip}
\frac{1}{\sqrt{20}} &   \frac{1}{\sqrt{20}}  &  \frac{1}{\sqrt{20}} &   \frac{1}{\sqrt{20}} &\frac{-4}{\sqrt{20}}  \\
\noalign{\smallskip}
\frac{1}{\sqrt{5}}  &    \frac{1}{\sqrt{5}}  &  \frac{1}{\sqrt{5}}  &   \frac{1}{\sqrt{5}}  &  \frac{1}{\sqrt{5}}   \\
\end{array}
\right )\left (
\begin{array}{cccccc}
\mathbf{r}_1   \\
\noalign{\smallskip}
\mathbf{r}_2    \\
\noalign{\smallskip}
\mathbf{r}_3      \\
\noalign{\smallskip}
\mathbf{r}_4       \\
\noalign{\smallskip}
\mathbf{r}_5    \\
\end{array}
\right ),
\end{eqnarray}
%\end{center}
%\end{widetext}
%\begin{widetext}
%\begin{center}
\begin{eqnarray}
\left (
\begin{array}{cccccc}
\mathbf{\zeta}_1   \\
\noalign{\smallskip}
\mathbf{\zeta}_2    \\
\noalign{\smallskip}
\mathbf{\zeta}_3      \\
\noalign{\smallskip}
\mathbf{\zeta}_4       \\
\noalign{\smallskip}
\mathbf{\zeta}_5    \\
\end{array}
\right )=\left (
\begin{array}{cccccc}
\frac{1}{\sqrt{2}}  &  \frac{-1}{\sqrt{2}}   &        0             &             0         &         0             \\
\noalign{\smallskip}
        0           &           0            &  \frac{-1}{\sqrt{2}} &   \frac{1}{\sqrt{2}}  &         0             \\
\noalign{\smallskip}
\frac{-17+\sqrt{241}}{2\sqrt{482-28\sqrt{241}}} & \frac{-17+\sqrt{241}}{2\sqrt{482-28\sqrt{241}}} & \frac{11-\sqrt{241}}{2\sqrt{482-28\sqrt{241}}} & \frac{11-\sqrt{241}}{2\sqrt{482-28\sqrt{241}}} & \frac{6}{\sqrt{482-28\sqrt{241}}}       \\
\noalign{\smallskip}
\frac{-17-\sqrt{241}}{2\sqrt{482+28\sqrt{241}}} &\frac{-17-\sqrt{241}}{2\sqrt{482+28\sqrt{241}}} & \frac{11+\sqrt{241}}{2\sqrt{482+28\sqrt{241}}} & \frac{11+\sqrt{241}}{2\sqrt{482+28\sqrt{241}}} & \frac{6}{\sqrt{482+28\sqrt{241}}}  \\
\noalign{\smallskip}
\frac{1}{\sqrt{5}}  &    \frac{1}{\sqrt{5}}  &  \frac{1}{\sqrt{5}}  &   \frac{1}{\sqrt{5}}  &  \frac{1}{\sqrt{5}}   \\
\end{array}
\right )\left (
\begin{array}{cccccc}
\mathbf{r}_1   \\
\noalign{\smallskip}
\mathbf{r}_2    \\
\noalign{\smallskip}
\mathbf{r}_3      \\
\noalign{\smallskip}
\mathbf{r}_4       \\
\noalign{\smallskip}
\mathbf{r}_5    \\
\end{array}
\right ), \\
\noalign{\smallskip}\noalign{\smallskip}
\left (
\begin{array}{cccccc}
\mathbf{\eta}_1   \\
\noalign{\smallskip}
\mathbf{\eta}_2    \\
\noalign{\smallskip}
\mathbf{\eta}_3      \\
\noalign{\smallskip}
\mathbf{\eta}_4       \\
\noalign{\smallskip}
\mathbf{\eta}_5    \\
\end{array}
\right )=\left (
\begin{array}{cccccc}
\frac{-3-\sqrt{5}}{5\sqrt{2}+\sqrt{10}}& \frac{-3-\sqrt{5}}{5\sqrt{2}+\sqrt{10}}  & \frac{-3-\sqrt{5}}{5\sqrt{2}+\sqrt{10}} &  \frac{-3-\sqrt{5}}{5\sqrt{2}+\sqrt{10}}  &  \frac{\sqrt{2}}{\sqrt{5}}  \\
\noalign{\smallskip}
\frac{3-\sqrt{5}}{5\sqrt{2}-\sqrt{10}} & \frac{2}{5\sqrt{2}-\sqrt{10}} & \frac{2}{5\sqrt{2}-\sqrt{10}} & \frac{3-\sqrt{5}}{5\sqrt{2}-\sqrt{10}} &\frac{-\sqrt{2}}{\sqrt{5}}     \\
\noalign{\smallskip}
\frac{-1}{2\sqrt{5+2\sqrt{5}}} & \frac{2+\sqrt{5}}{2\sqrt{5+2\sqrt{5}}} & \frac{-2-\sqrt{5}}{2\sqrt{5+2\sqrt{5}}} & \frac{1}{2\sqrt{5+2\sqrt{5}}} & 0  \\
\noalign{\smallskip}
\frac{-1}{2\sqrt{5-2\sqrt{5}}} & \frac{2-\sqrt{5}}{2\sqrt{5-2\sqrt{5}}} & \frac{-2+\sqrt{5}}{2\sqrt{5-2\sqrt{5}}} & \frac{1}{2\sqrt{5-2\sqrt{5}}} & 0 \\
\frac{1}{\sqrt{5}}  &    \frac{1}{\sqrt{5}}  &  \frac{1}{\sqrt{5}}  &   \frac{1}{\sqrt{5}}  &  \frac{1}{\sqrt{5}}   \\
\end{array}
\right ) \left (
\begin{array}{cccccc}
\mathbf{r}_1   \\
\noalign{\smallskip}
\mathbf{r}_2    \\
\noalign{\smallskip}
\mathbf{r}_3      \\
\noalign{\smallskip}
\mathbf{r}_4       \\
\noalign{\smallskip}
\mathbf{r}_5    \\
\end{array}
\right ).~~~~~~~~~~~~~~~~~
\end{eqnarray}
\end{center}
\end{widetext}
Finally, we simplify the minimums of those quadratic confinement potentials into the sum 
of several independent harmonic oscillators,
\begin{eqnarray}
\begin{array}{cccccccccc}
V_{1min}^{con}(5)=k\left(\xi_1^2+\xi_2^2+\xi_3^2\right), \hspace{3.65cm}
\end{array} \\
\begin{array}{cccccccccc}
\noalign{\smallskip}
V_{2min}^{con}(5)=k\left(\chi_1^2+\chi_2^2+\frac{1}{3}\chi_3^2+\frac{5}{7}\chi_4^2\right),\hspace{2.25cm}
\end{array} \\
\begin{array}{cccccccccc}
V_{3min}^{con}(5)=k\left(\zeta_1^2+\zeta_2^2+\frac{46+\sqrt{241}}{75}\zeta_3^2+\frac{46-\sqrt{241}}{75}\zeta_4^2\right),\hspace{0.3cm} \\
\end{array} \\
\begin{array}{cccccccccc}
V_{4min}^{con}(5)=k\left(\frac{15+\sqrt{5}}{22}(\eta_1^2+\eta_3^2)+\frac{15-\sqrt{5}}{22}(\eta_2^2+\eta_4^2)\right). \hspace{0.3cm}
\end{array}
\end{eqnarray}

The perturbative effect of QCD can be described by the one-gluon-exchange (OGE) interaction.
From the non-relativistic reduction of the OGE diagram in QCD for point-like quarks one gets
\begin{eqnarray}
V_{ij}^{oge} & = & {\frac{\alpha_{s}}{4}}\boldsymbol{\lambda}^c_{i}\cdot\boldsymbol{\lambda}_{j}^c\left({\frac{1}{r_{ij}}}-
{\frac{2\pi\delta(\mathbf{r}_{ij})\boldsymbol{\sigma}_{i}\cdot
\boldsymbol{\sigma}_{j}}{3m_im_j}}\right),
\end{eqnarray}
$m_i$ is the effective mass of the $i$-th quark. $\mathbf{r}_{ij}=\mathbf{r}_i-\mathbf{r}_j$ and
$r_{ij}=|\mathbf{r}_{i}-\mathbf{r}_{j}|$. $\boldsymbol{\lambda}^c$ and $\boldsymbol{\sigma}$ represent
the Gell-Mann matrices and the Pauli matrices, respectively. Dirac $\delta(\mathbf{r}_{ij})$ function
comes out in the deduction of the interaction between point-like quarks, when not treated perturbatively,
which leads to collapse~\cite{collapse}. Therefore, the $\delta(\mathbf{r}_{ij})$ function can be
regularized in the form~\cite{vijande}
\begin{equation}
\delta(\mathbf{r}_{ij})\rightarrow\frac{1}{4\pi r_{ij}r_0^2(\mu_{ij})}e^{-\frac{r_{ij}}{r_0(\mu_{ij})}},
\end{equation}
where $r_0(\mu_{ij})=\frac{r_0}{\mu_{ij}}$, in which $r_0$ is an adjustable model parameter and $\mu_{ij}$ is
the reduced mass of two interacting particles. This regularization is justified based on the finite size of the
constituent quarks and should be therefore flavor dependent~\cite{flavor-dependent}.

The quark-gluon coupling constant takes an effective scale-dependent form,
\begin{equation}
\alpha_s(\mu^2_{ij})=\frac{\alpha_0}{\ln\frac{\mu_{ij}^2}{\Lambda_0^2}},
\end{equation}
$\Lambda_0$ and $\alpha_0$ are adjustable model parameters.

To sum up, the completely Hamiltonian of the MCFTM for the mesons, baryons and hidden charm
pentaquark states can be presented as
\begin{eqnarray}
H_n =\sum_{i=1}^n \left(m_i+\frac{\mathbf{p}_i^2}{2m_i}\right)-T_{c}+\sum_{i<j}^n V_{ij}^{oge}+V_{min}^{con}(n).
\end{eqnarray}
$T_{c}$ is the center-of-mass kinetic energy and should be deducted; $\mathbf{p}_i$
is the momentum of the $i$-th quark.

\section{wavefunctions}

The total wavefunction $\Phi^{P_{cs}}_{IJ}$ of the pentaquark ground state $[nn][cs]\bar{c}$ ($n=u$ and $d$)
with well-defined isospin $I$ and angular momentum $J$ reads
\begin{eqnarray}
\Phi^{P_{cs}}_{IJ}=\sum_{\delta}c_{\delta}\left[\left[\Psi_{csi}^{[nn]}\Psi_{csi}^{[cs]}\Psi_{csi}^{\bar{c}}
\right]_{IS}F(\mathbf{r},\mathbf{R},\boldsymbol{\lambda},\boldsymbol{\rho})\right]_{IJ}, \nonumber\\
\end{eqnarray}
where all [~]s represent all possible Clebsch-Gordan (C-G) coupling. $\Psi_{csi}$s are the color-spin-isospin
($csi$) wave functions and can be written as the product of the wave functions of color $\psi_c$, isospin $\omega_i$
and spin $\chi_s$,
\begin{eqnarray}
\Psi_{csi}^{[nn]}&=&\psi_{c}^{[nn]}\chi_{ss_z}^{[nn]}\omega_{ii_z}^{[nn]},~
\Psi_{csi}^{[cs]}=\psi_{c}^{[cs]}\chi_{ss_z}^{[cs]}\omega_{ii_z}^{[cs]},\\
 \noalign{\smallskip}
\Psi_{csi}^{\bar{c}}&=&\psi_{c}^{\bar{c}}\chi_{ss_z}^{\bar{c}}\omega_{ii_z}^{\bar{c}}.
\end{eqnarray}

A set of Jacobi coordinates $\mathbf{r}$, $\mathbf{R}$, $\boldsymbol{\lambda}$, and $\boldsymbol{\rho}$
are used to describe the relative motions in the state $P_{cs}$,
\begin{eqnarray}
\mathbf{r}&=&\mathbf{r}_1-\mathbf{r}_2,~\mathbf{R}~=~\mathbf{r}_3-\mathbf{r}_4,~
\boldsymbol{\lambda}~=~\frac{\mathbf{r}_1+\mathbf{r}_2}{2}-\mathbf{r}_5,~~~\\
\boldsymbol{\rho}&=&\frac{m_{n}\mathbf{r}_1+m_n\mathbf{r}_2+m_c\mathbf{r}_5}{2m_n+m_c}
-\frac{m_c\mathbf{r}_3+m_s\mathbf{r}_4}{m_s+m_c}.
\end{eqnarray}
Only the ground states are investigated in this work. The total spatial wave function
$F(\mathbf{r},\mathbf{R},\boldsymbol{\lambda},\boldsymbol{\rho})$ can be separated into a
product of four relative motion wave functions
\begin{eqnarray}
F(\mathbf{r},\mathbf{R},\boldsymbol{\lambda},\boldsymbol{\rho})=\phi_{00}(\mathbf{r})
\phi_{00}(\mathbf{R})\phi_{00}(\boldsymbol{\lambda})\phi_{00}(\boldsymbol{\rho}).
\end{eqnarray}

According to the Gaussian expansion method (GEM) \cite{GEM}, the relative motion wave
function $\phi_{lm}(\mathbf{x})$, where $\mathbf{x}$ stands for $\mathbf{r}$, $\mathbf{R}$,
$\boldsymbol{\lambda}$, and $\boldsymbol{\rho}$, can be expanded as the superposition
of many different size ($\nu_n$) Gaussian functions with well-defined orbital angular
momentum,
\begin{eqnarray}
\phi_{lm}(\mathbf{x})=\sum_{n=1}^{n_{max}}c_{n}N_{nl}x^{l}e^{-\nu_{n}x^2}Y_{lm}(\hat{\mathbf{x}})
\end{eqnarray}
Gaussian size parameters are taken as geometric progression,
\begin{eqnarray}
\nu_{n}=\frac{1}{r^2_n}, &r_n=r_1a^{n-1},
&a=\left(\frac{r_{n_{max}}}{r_1}\right)^{\frac{1}{n_{max}-1}}
\end{eqnarray}
$N_{nl}$ is normalized coefficient and $c_n$ is a variation coefficient determined
by the model dynamics. With $r_1=0.2$ fm, $r_{n_{max}}=2.0$ fm and $n_{max}=7$, the
converged numerical results can be achieved in the present work.

The spin wave functions $\chi^{[nn]}_{ss_z}$ of the diquark $[nn]$ can be written as
\begin{eqnarray}
\chi_{10}^{[nn]}&=&\frac{1}{\sqrt{2}}(\uparrow\downarrow+\downarrow\uparrow),~\chi_{11}^{[nn]}=\uparrow\uparrow,\\
\chi_{1-1}^{[nn]}&=&\downarrow\downarrow,~\chi_{00}^{[nn]}=\frac{1}{\sqrt{2}}(\uparrow\downarrow-\downarrow\uparrow).
\end{eqnarray}
where $\uparrow$ and $\downarrow$ represent spin up and spin down, respectively. The
wave functions $\chi_{ss_z}^{[cs]}$ of the diquark $[cs]$ are exactly same with
$\chi_{ss_z}^{[nn]}$. The wave functions $\chi_{ss_z}^{[\bar{c}]}$ of the antiqark
$\bar{c}$ read
\begin{eqnarray}
\chi_{\frac{1}{2}\frac{1}{2}}^{\bar{c}}=\uparrow,~\chi_{\frac{1}{2}-\frac{1}{2}}^{\bar{c}}=\downarrow.
\end{eqnarray}
The total spin wave function of the state $P_{cs}$ with spin $S$ and $z$-component $S_z$
can be obtained by the following Clebsch-Gordan coupling
\begin{eqnarray}
\chi_{SS_z}^{P_{cs}}&=\chi^{[nn]}_{ss_z}\oplus\chi^{[cs]}_{ss_z}\oplus\chi^{\bar{c}}_{ss_z},
\end{eqnarray}

The isospin wave functions $\omega_{ii_z}^{[nn]}$, $\omega_{ii_z}^{[cs]}$ and $\omega_{ii_z}^{\bar{c}}$
can be expressed as
\begin{eqnarray}
\omega_{10}^{[nn]}&=&\frac{1}{\sqrt{2}}(ud+du),~\omega_{11}^{[nn]}=uu,~\omega_{1-1}^{[nn]}=dd,\\
\omega_{00}^{[nn]}&=&\frac{1}{\sqrt{2}}(ud-du),~\omega_{00}^{[cs]}=cs,~\omega_{00}^{\bar{c}}=\bar{c}.
\end{eqnarray}
The isospin of the state $P_{cs}$ is determined by the diquark $[nn]$ because those of the diquarks
$[cs]$ and $\bar{c}$ are zero. The total isospin wave function therefore reads
\begin{eqnarray}
\omega_{II_z}^{P_{cs}}&=&\omega^{[nn]}_{II_z}\omega^{[cs]}_{00}\omega^{\bar{c}}_{00},
\end{eqnarray}

The color wave functions of the diquark $[nn]$ can be antisymmetrical color $\bar{\mathbf{3}}$
and symmetrical color $\mathbf{6}$ representation, their explicit component expressions read
\begin{eqnarray}
\psi_{\bar{\mathbf{3}}_1}^{[nn]}&=&\frac{1}{\sqrt{2}}\left(rg-gr\right),
~\psi_{\bar{\mathbf{3}}_2}^{[nn]}=\frac{1}{\sqrt{2}}(gb-bg), \\
\psi_{\bar{\mathbf{3}}_3}^{[nn]}&=&\frac{1}{\sqrt{2}}(br-rb),~\psi_{\mathbf{6}_1}^{[nn]}=rr,\\
\psi_{\mathbf{6}_2}^{[nn]}&=&\frac{1}{\sqrt{2}}(rg+gr),~\psi_{\mathbf{6}_3}^{[nn]}=\frac{1}{\sqrt{2}}(rb+br),\\ \psi_{\mathbf{6}_4}^{[nn]}&=&gg,~\psi_{\mathbf{6}_5}^{[nn]}=\frac{1}{\sqrt{2}}(gb+bg),~\psi_{\mathbf{6}_6}^{[nn]}=bb
\end{eqnarray}
Those of the diquark $[cs]$ are exactly same with the diquark $[nn]$. The antiquark $\bar{c}$
is in color $\bar{\mathbf{3}}$ and read
\begin{eqnarray}
\psi_{\bar{\mathbf{3}}_1}^{\bar{c}}&=&\bar{r},~\psi_{\bar{\mathbf{3}}_2}^{\bar{c}}=\bar{g},
~\psi_{\bar{\mathbf{3}}_3}^{\bar{c}}=\bar{b}.
\end{eqnarray}
The diquarks $\psi_{c}^{[nn]}$ and $\psi_{c}^{[cs]}$ must couple into a tetraquark state in
color $\mathbf{3}$ according to the requirement of overall color singlet of the state $P_{cs}$.
Therefore, the total color singlet wave function can be expressed as
\begin{eqnarray}
\psi_c^{P_{cs}}=\frac{1}{\sqrt3}\left(\psi_{\mathbf{3}_1}^{[nn][cs]}
\psi_{\bar{\mathbf{3}}_1}^{\bar{c}}+\psi_{\mathbf{3}_2}^{[nn][cs]}\psi_{\bar{\mathbf{3}}_2}^{\bar{c}}
+\psi_{\mathbf{3}_3}^{[nn][cs]}\psi_{\bar{\mathbf{3}}_3}^{\bar{c}}\right).\nonumber\\
\end{eqnarray}

There are the following three different coupling ways of the diquark $[nn]$ and $[cs]$ into
a tetraquark state $\psi_{\mathbf{3}}^{[nn][cs]}$,
case A: $\psi_{\bar{\mathbf{3}}}^{[nn]}\otimes\psi_{\bar{\mathbf{3}}}^{[cs]}$;
case B: $\psi_{\mathbf{6}}^{[nn]}\otimes\psi_{\bar{\mathbf{3}}}^{[cs]}$ and
case C: $\psi_{\bar{\mathbf{3}}}^{[nn]}\otimes\psi_{\mathbf{6}}^{[cs]}$.
For the case A, its explicit component expressions read
\begin{eqnarray}
\psi_{\mathbf{3}_1}^{[nn][cs]}=\frac{1}{\sqrt2}\psi_{\bar{\mathbf{3}}_1}^{[nn]}\psi_{\bar{\mathbf{3}}_3}^{[cs]}
-\frac{1}{\sqrt2}\psi_{\bar{\mathbf{3}}_3}^{[nn]}\psi_{\bar{\mathbf{3}}_1}^{[cs]},\\
\psi_{\mathbf{3}_2}^{[nn][cs]}=\frac{1}{\sqrt2}\psi_{\bar{\mathbf{3}}_1}^{[nn]}\psi_{\bar{\mathbf{3}}_2}^{[cs]}
-\frac{1}{\sqrt2}\psi_{\bar{\mathbf{3}}_2}^{[nn]}\psi_{\bar{\mathbf{3}}_1}^{[cs]},\\
\psi_{\mathbf{3}_3}^{[nn][cs]}=\frac{1}{\sqrt2}\psi_{\bar{\mathbf{3}}_3}^{[nn]}\psi_{\bar{\mathbf{3}}_2}^{[cs]}
-\frac{1}{\sqrt2}\psi_{\bar{\mathbf{3}}_2}^{[nn]}\psi_{\bar{\mathbf{3}}_3}^{[cs]}.
\end{eqnarray}
For the case B, its explicit component expressions read
\begin{eqnarray}
\psi_{\mathbf{3}_1}^{[nn][cs]}=\frac{\sqrt{2}}{2}\psi^{[nn]}_{\mathbf{6}_1}
\psi_{\bar{\mathbf{3}}_2}^{[cs]}-\frac{1}{2}\psi^{[nn]}_{\mathbf{6}_2}\psi_{\bar{\mathbf{3}}_3}^{[cs]}+
\frac{1}{2}\psi^{[nn]}_{\mathbf{6}_3}\psi_{\bar{\mathbf{3}}_1}^{[cs]},\nonumber\\ \\
\psi_{\mathbf{3}_2}^{[nn][cs]}=\frac{1}{2}\psi^{[nn]}_{\mathbf{6}_2}
\psi_{\bar{\mathbf{3}}_2}^{[cs]}-\frac{\sqrt{2}}{2}\psi^{[nn]}_{\mathbf{6}_4}\psi_{\bar{\mathbf{3}}_3}^{[cs]}+
\frac{1}{2}\psi^{[nn]}_{\mathbf{6}_5}\psi_{\bar{\mathbf{3}}_1}^{[cs]},\nonumber\\ \\
\psi_{\mathbf{3}_3}^{[nn][cs]}=\frac{1}{2}\psi^{[nn]}_{\mathbf{6}_3}
\psi_{\bar{\mathbf{3}}_2}^{[cs]}-\frac{1}{2}\psi^{[nn]}_{\mathbf{6}_5}\psi_{\bar{\mathbf{3}}_3}^{[cs]}
+\frac{\sqrt{2}}{2}\psi^{[nn]}_{\mathbf{6}_6}\psi_{\bar{\mathbf{3}}_1}^{[cs]}.\nonumber\\
\end{eqnarray}
For the case C, its explicit component expressions read
\begin{eqnarray}
\psi_{\mathbf{3}_1}^{[nn][cs]}=\frac{\sqrt{2}}{2}\psi_{\bar{\mathbf{3}}_2}^{[nn]}
\psi^{[cs]}_{\mathbf{6}_1}-\frac{1}{2}\psi_{\bar{\mathbf{3}}_3}^{[nn]}\psi^{[cs]}_{\mathbf{6}_2}+
\frac{1}{2}\psi_{\bar{\mathbf{3}}_1}^{[nn]}\psi^{[cs]}_{\mathbf{6}_3},\nonumber\\ \\
\psi_{\mathbf{3}_2}^{[nn][cs]}=\frac{1}{2}\psi_{\bar{\mathbf{3}}_2}^{[nn]}
\psi^{[cs]}_{\mathbf{6}_2}-\frac{\sqrt{2}}{2}\psi_{\bar{\mathbf{3}}_3}^{[nn]}\psi^{[cs]}_{\mathbf{6}_4}+
\frac{1}{2}\psi_{\bar{\mathbf{3}}_1}^{[nn]}\psi^{[cs]}_{\mathbf{6}_5},\nonumber\\ \\
\psi_{\mathbf{3}_3}^{[nn][cs]}=\frac{1}{2}\psi_{\bar{\mathbf{3}}_2}^{[nn]}
\psi^{[cs]}_{\mathbf{6}_3}-\frac{1}{2}\psi_{\bar{\mathbf{3}}_3}^{[nn]}\psi^{[cs]}_{\mathbf{6}_5}
+\frac{\sqrt{2}}{2}\psi_{\bar{\mathbf{3}}_1}^{[nn]}\psi^{[cs]}_{\mathbf{6}_6}.\nonumber\\
\end{eqnarray}

The diquark is a spatially extended object with various color-spin-isospin-orbit
combinations~\cite{diquark}. For the sake of convenience, we define the color quantum
number $c=0$ and $c=1$ for the diquark in the color $\bar{\mathbf{3}}$ and $\mathbf{6}$
representation, respectively. For the identical diquarks $[dd]$, $[ud]$ and $[ud]$,
their spin $s$, isospin $i$, orbit angular excitation $l$, and color $c$ obey the
constraint $s+i+l+c=even$ to satisfy the Pauli principle. The spin singlet, isospin
singlet and color triplet diquark with $l=0$ is often called the good diquark. Other
combinations are sometimes called bad diquarks. For the strange diquark $[ss]$, its
spin $s$, isospin $i$, orbit angular excitation $l$, and color $c$ obey the constraint
$s+i+l+c=odd$ because its isospin is symmetrical. In this work, we are only
interested in the ground states, namely $l=0$. The diquarks $[cu]$, $[cd]$
and $[cs]$ are not identical particles so that their quantum numbers are not
constrained.

The diquark $[nn]$ has four possible antisymmetrical spin-isospin-color combinations,
\begin{eqnarray}
\Psi_{csi}^{[nn]}=\Psi_{\bar{\mathbf{3}}00}^{[nn]},~\Psi_{\bar{\mathbf{3}}11}^{[nn]},
~\Psi_{\mathbf{6}01}^{[nn]},~\Psi_{\mathbf{6}10}^{[nn]}.
\end{eqnarray}
According to the total spin and isospin of the state $P_{cs}$ and color configurations,
one can obtain all possible wave functions, which are represented by the $\delta$ in Eq. 
(20). Its corresponding coefficient $c_{\delta}$ can be determined by the model dynamics. 
For example, the total wave function of the state $P_{cs}$ with $0\frac{1}{2}^-$ has seven
possibilities.

The total wave function $\Phi^{P_{css}}_{IJ}$, $P_{css}=[ss][cn]\bar{c}$, is exactly
same with that of the state $P_{cs}$ with isospin $I=1$ because the flavor parts
of the diquarks $[ss]$ and $[nn]$ are both symmetrical. For the same reason, the total
wave function $\Phi^{P_{csss}}_{IJ}$, $P_{csss}=[ss][cs]\bar{c}$, is exactly same with
that of the state $[nn][cn]\bar{c}$ with isospin $I=\frac{3}{2}$, which can be obtained
in the previous work~\cite{4380cftm}.

Using the same procedure with the diquark configuration, one can easily construct the 
wave functions of the state $[qqq][c\bar{c}]$ with color octet configuration, which can 
also be achieved in ref.~\cite{wave8} so that those are omitted here. Note that it is
difficult to construct the wave functions of the pentagon structure in the quark level.
In this work, we first employ the wave functions of the diquark structure to calculate 
the mass of the pentagon structure approximately. More reliable estimation is left for
further research in future.  

\section{numerical results and discussions}

\subsection{Parameters and conventional hadron spectra }

We take the $\phi$ and $\omega$ mesons as the ideal mixing of the SU(3) singlet $\omega_0$ 
and the octet $\omega_8$ states in this work, namely $\omega=\frac{1}{\sqrt{2}}(u\bar{u}+d\bar{d})$,
$\phi=s\bar{s}$ and the ideal mixing angle $\theta_V=35.3^o$. We can obtain the masses 
of meson and baryon ground states by approximately strict solving two-body and three-body 
Schr\"{o}dinger equations in the MCFTM. We use the mean square error
\begin{eqnarray}
\Delta=\sum_{i=1}^N\frac{w_i(M_i-m_i)^2}{N}
\end{eqnarray}
to fit the mass spectra and to determine the adjustable parameters and their errors
in the Minuit program~\cite{4380cftm}. $N$ is the total number of mesons and baryons.
$M_i$ is the experimental mass of the ith meson or baryon and $m_i$ is its predicted 
mass in the model. $w_i$ is its corresponding weight for fitting mass spectrum better. 
For the heavy parts, their weights are equal to 1. For the light parts, especially for 
$\pi$ and $K$ mesons, their values are greater than 1, such as 2 and 3.

Finally, we can obtain the optima parameters and spectra, which are presented in 
Tables~\ref{parameters} and \ref{meson-baryon}, respectively. Moreover, we can also arrive 
at the mass errors for mesons, baryons and pentaquark states, just several MeV~\cite{4380cftm}, 
introduced by the errors of the parameters. From Table \ref{meson-baryon}, one can 
see that the meson and baryon ground states, from the lightest $\pi$ to the heaviest 
$\Upsilon(1S)$, can be simultaneously accommodated in the model very well with only a few 
adjustable model parameters. The fact indicates that the multibody confinement potential 
based on the color flux-tube picture may be a valid dynamical mechanism in the phenomenological
description of the properties of meson and baryon states. Of course, other properties
of those states need further study, which is left for the future work.
\begin{table}[ht]
\caption{Adjustable model parameters, quark mass and $\Lambda_0$ unit in MeV,
$k$ unit in MeV$\cdot$fm$^{-2}$, $r_0$ unit in MeV$\cdot$fm and $\alpha_0$ is
dimensionless.}\label{parameters}
\begin{tabular}{ccccccccccc}
\toprule[0.8pt] \noalign{\smallskip}
Para.   & $m_{u,d}$  & ~~$m_{s}$~~ & ~~$m_c$~~  & ~~$m_b$~~    &  ~~$k$~~  &  ~~~$\alpha_0$~~  & ~~~$\Lambda_0$~~~  & ~~$r_0$~~  \\
Valu.   &     230    &     473     &   1701     &    5047      &     700   &       4.69        &         30.24      &   81.48    \\
\toprule[0.8pt] \noalign{\smallskip}
\end{tabular}
%\end{table}
%\begin{table}[ht]
\caption{Conventional meson and baryon spectra, unit in MeV.\label{meson-baryon}}
\begin{tabular}{cccccccccccccccccccccccc}
\toprule[0.8pt] \noalign{\smallskip}
State  &&       $\pi$    &    $\rho$  &    $\omega$  &    $K$     &   $K^*$     &     $\phi$    &     $D^{\pm}$   \\
\noalign{\smallskip}
Theo.  &&       137      &    762     &     762      &    494     &     922     &      1058     &      1879       \\
PDG    &&       139      &    770     &     780      &    496     &     896     &      1020     &      1869       \\
\toprule[0.8pt] \noalign{\smallskip}
State  &&      $D^*$     & $D_s^{\pm}$&    $D_s^*$   &  $\eta_c$  &   $J/\Psi$  &     $B^0$     &      $B^*$      \\
\noalign{\smallskip}
Theo.  &&       2039     &    1952    &     2144     &    2949    &     3128    &      5285     &      5343       \\
PDG    &&       2007     &    1968    &     2112     &    2980    &     3097    &      5280     &      5325       \\
\toprule[0.8pt] \noalign{\smallskip}
State  &&      $B_s^0$   &    $B_s^*$ &     $B_c$    &   $B_c^*$  &   $\eta_b$  &   $\Upsilon(1S)$                \\
\noalign{\smallskip}
Theo.  &&       5352     &    5429    &     6254     &    6396    &     9374    &      9536                       \\
PDG    &&       5366     &    5416    &     6277     &    ...     &     9391    &      9460                       \\
\toprule[0.8pt] \noalign{\smallskip}
State  &&       $N$      &  $\Delta$  &    $\Sigma$  & $\Sigma^*$ &     $\Xi$   &     $\Xi^*$   &   $\Lambda$     \\
\noalign{\smallskip}
Theo.  &&       945      &    1239    &     1204     &    1391    &     1345    &      1537     &      1128       \\
PDG    &&       939      &    1232    &     1195     &    1385    &     1315    &      1530     &      1115       \\
\toprule[0.8pt] \noalign{\smallskip}
State  &&     $\Omega$   & $\Sigma_c$ & $\Sigma_c^*$ &   $\Xi_c$  &$   \Xi_c^*$ &  $\Omega_c^0$ & $\Omega_c^{0*}$ \\
\noalign{\smallskip}
Theo.  &&       1677     &    2437    &     2508     &    2460    &     2626    &      2703     &      2774       \\
PDG    &&       1672     &    2445    &     2520     &    2466    &     2645    &      2695     &      2766       \\
\toprule[0.8pt] \noalign{\smallskip}
State  &&  $\Lambda_c^+$ & $\Sigma_b$ & $\Sigma_b^*$ &  $\Xi_b$   &  $\Xi_b^*$  &  $\Omega_b^-$ & $\Lambda_b^0$   \\
\noalign{\smallskip}
Theo.  &&       2278     & ~~~5786~~~ &     5812     & ~~~5765~~~ &     5817    &   ~~~6034~~~  &      5596       \\
PDG    &&       2285     &    5808    &     5830     &    5790    &     ...     &      6071     &      5620       \\
\toprule[0.8pt] \noalign{\smallskip}
\end{tabular}
\end{table}

\subsection{$qqqc\bar{c}$ spectrum}

So far, the baryon-meson molecular descriptions of the $P_c$ and $P_{cs}$ states seem to 
prevail over other possibilities in various theoretical framework because of the proximity
of their masses to the baryon-meson thresholds~\cite{reviews}. However, it does not 
mean that other possibilities can be excluded completely. According to QCD, the hidden
color components are allowed in addition to the color singlet component in the pentaquark
states. In a large extent, the pentaquark states should be a mixture of all possible
color configurations. In this work, we first attempt to explore the natures of the
pentaquark states from the perspective of hidden color components. Another reason
is the absence of the one-boson-exchange interaction in the MCFTM, which is widely
accepted as the binding mechanism of molecular states from the phenomenological model
point of view. The mixing between the color singlet and hidden color components
deserves further investigation in future.

Next, we move on to the investigation on the properties of the hidden color 
pentaquark states $qqqc\bar{c}$ in the MCFTM. The $P$-parity of the states is 
negative because we are interested in the ground states. In this way, the 
spin-parity assignment of the pentaquark states should be $\frac{1}{2}^-$, 
$\frac{3}{2}^-$ and $\frac{5}{2}^-$. The total isospin of the pentaquark states 
depends on their specific quark content. We can achieve the mass of the states 
with all possible isospin and spin-parity and three various color structures, 
diquark, octet and pentagon, by solving the five-body Schr\"{o}dinger equation 
with the well-defined trial wave functions. We present their mass spectrum in 
Table~\ref{hiddencharm}, in which $E_d$, $E_o$, and $E_p$ respectively represent 
the masses of the diquark, octet and pentagon structures.

\begin{table}[ht]
\caption{The mass of the ground state $qqqc\bar{c}$ with $IJ^P$ and various color structures, unit in MeV.\label{hiddencharm}}
\begin{tabular}{cccccccccccccccccccc}
\toprule[0.8pt] \noalign{\smallskip}
&~~~$nnnc\bar{c}$, $I=\frac{1}{2}$~~~&&~~~$nnnc\bar{c}$, $I=\frac{3}{2}$~~~&&~~~$nnsc\bar{c}$, $I=0$~~~\\
\noalign{\smallskip}
$J^P$ &~$E_o$~~~~$E_d$~~~~$E_p$~ && ~$E_o$~~~~$E_d$~~~~$E_p$~ && ~$E_o$~~~~$E_d$~~~~$E_p$~ \\
\noalign{\smallskip}\toprule[0.8pt] \noalign{\smallskip}
$\frac{1}{2}^-$ & 4402~~4344~~4303  && 4620~~4581~~4532 && 4512~~4487~~4463  \\
$\frac{3}{2}^-$ & 4473~~4405~~4369  && 4661~~4622~~4573 && 4611~~4585~~4570  \\
$\frac{5}{2}^-$ & 4616~~4569~~4516  && 4743~~4707~~4666 && 4911~~4884~~4846  \\
\noalign{\smallskip}\toprule[0.8pt] \noalign{\smallskip}
&~~~$nnsc\bar{c}$, $I=1$~~~&&~~~$nssc\bar{c}$, $I=\frac{1}{2}$~~~&&~~~$sssc\bar{c}$, $I=0$~~~\\
\noalign{\smallskip}
$J^P$ &~$E_o$~~~~$E_d$~~~~$E_p$~ && ~$E_o$~~~~$E_d$~~~~$E_p$~ && ~$E_o$~~~~$E_d$~~~~$E_p$~ \\
\noalign{\smallskip}\toprule[0.8pt] \noalign{\smallskip}
$\frac{1}{2}^-$ & 4617~~4595~~4579 && 4784~~4750~~4730 && ~5047~~5019~~4985\\
$\frac{3}{2}^-$ & 4715~~4690~~4675 && 4877~~4839~~4823 && ~5074~~5048~~5017\\
$\frac{5}{2}^-$ & 4850~~4822~~4810 && 5008~~4963~~4954 && ~5140~~5115~~5089\\
\noalign{\smallskip}\toprule[0.8pt] \noalign{\smallskip}
\end{tabular}
\caption{The average values $\langle V^{oge}\rangle$, $\langle V^{con}\rangle$, and $\langle T \rangle$
in the three structures, $T$ stands for kinetic energy, unit in MeV.\label{averagevalue}}
\begin{tabular}{ccccccccccc}
\toprule[0.8pt] \noalign{\smallskip}
Flavor&&$nnnc\bar{c}$ & & &$nnsc\bar{c}$&  \\
$IJ^P$&&$\frac{1}{2}\frac{1}{2}^-$ && &$0\frac{1}{2}^-$& \\
\noalign{\smallskip}
Stru. & octet&diquark&pentagon &octet&diquark&pentagon\\
\noalign{\smallskip}
$\langle V^{oge}\rangle$ & $-2111$  & $-2099$    & $-2074$  & $-1999$  & $-1973$    & $-1965$  \\
$\langle V^{con}\rangle$ & $~~1756$ & $~~~1710~$ & $~~1673$ & $~~1628$ & $~~~1594~$ & $~~1574$ \\
$\langle T \rangle$      & $~664$   & $~640$     & $~614$   & $~547$   & $~529$     & $~518$   \\
\toprule[0.8pt] \noalign{\smallskip}
\end{tabular}
\end{table}

It can be seen from Table~\ref{hiddencharm} that the color structures can induce the mass
splitting like the color-magnetic interaction does. The masses $E_o$, $E_d$ and $E_p$ are
close and their order is $E_o>E_d>E_p$. The mass difference between the two adjacent items
is several tens MeV, which mainly come from the different type of confinement potential
determined by the color structure, see Table~\ref{averagevalue}. The confinement
potential of the octet structure is bigger than that of the diquark structure because there
is one piece of stronger color $\mathbf{8}$-dimension color flux-tube than $\mathbf{3}$-dimension
one. That of the ring-like pentagon structure is lowest because the structure is easier to shrink
into a compact multiquark state relative to the octet and diquark structures.

\subsection{$P_c$ and $P_{cs}$ states observed by the LHCb Collaboration}

Matching the masses predicted by the MCFTM with the experimental data of the states, we present
the possible interpretation on the $IJ^P$ and color structures of the states in Table~\ref{ours}.
At first glance, all of the states can be accommodated in the model. In addition, we also calculate 
the average distances, smaller than 1 fm, between any two quarks using the eigen wave function of 
the states. In this way, the states should be compact in the model because of the five-body confinement 
potential.

\begin{table*}[ht]
\centering
\caption{Possible isospin-spin-parity and structure assignments of the $P_c$ and $P_{cs}$ states and 
their average distance $\langle\mathbf{r}_{ij}^2\rangle^{\frac{1}{2}}$ in the MCFTM , unit in fm.\label{ours}}
\begin{tabular}{ccccccccccccccc}
\toprule[0.8pt] \noalign{\smallskip}
Flavor&$IJ^P$&Structure&Mass&~Candidate~&$\langle\mathbf{r}_{12}^2\rangle^{\frac{1}{2}}$~&~$\langle\mathbf{r}_{13}^2\rangle^{\frac{1}{2}}$~ &~$\langle\mathbf{r}_{23}^2\rangle^{\frac{1}{2}}$~& ~$\langle\mathbf{r}_{14}^2\rangle^{\frac{1}{2}}$~ &~$\langle\mathbf{r}_{24}^2\rangle^{\frac{1}{2}}$~& ~$\langle\mathbf{r}_{34}^2\rangle^{\frac{1}{2}}$~& ~$\langle\mathbf{r}_{15}^2\rangle^{\frac{1}{2}}$~ &~$\langle\mathbf{r}_{25}^2\rangle^{\frac{1}{2}}$~& ~$\langle\mathbf{r}_{35}^2\rangle^{\frac{1}{2}}$~&~$\langle\mathbf{r}_{45}^2\rangle^{\frac{1}{2}}$~    \\
\toprule[0.8pt]
\noalign{\smallskip}
$uudc\bar{c}$&$\frac{1}{2}\frac{1}{2}^-$&pentagon&4303&$P_c(4312)^+$    &0.90 & 0.90 & 0.90 & 0.75 & 0.75 & 0.75 & 0.76 & 0.76 & 0.76 & 0.37  \\
\noalign{\smallskip}%\toprule[0.8pt]
$uudc\bar{c}$&$\frac{1}{2}\frac{1}{2}^-$& diquark&4344&$P_c(4337)^+$    &0.90 & 0.90 & 0.90 & 0.75 & 0.75 & 0.75 & 0.76 & 0.76 & 0.76 & 0.37  \\
\noalign{\smallskip}%\toprule[0.8pt]
$uudc\bar{c}$&$\frac{1}{2}\frac{3}{2}^-$&pentagon&4369&$P_c(4380)^+$    &0.91 & 0.91 & 0.91 & 0.78 & 0.78 & 0.78 & 0.78 & 0.78 & 0.78 & 0.40  \\
\noalign{\smallskip}%\toprule[0.8pt]
$uudc\bar{c}$&$\frac{1}{2}\frac{3}{2}^-$& diquark&4405&$P_c(4440)^+$    &0.89 & 0.89 & 0.89 & 0.77 & 0.77 & 0.77 & 0.77 & 0.77 & 0.77 & 0.40  \\
\noalign{\smallskip}%\toprule[0.8pt]
$uudc\bar{c}$&$\frac{1}{2}\frac{3}{2}^-$&  octet &4475&$P_c(4457)^+$    &0.89 & 0.89 & 0.89 & 0.77 & 0.77 & 0.77 & 0.77 & 0.77 & 0.77 & 0.40  \\
\noalign{\smallskip}%\toprule[0.8pt]
$udsc\bar{c}$&$0\frac{1}{2}^-$          &pentagon&4463&$P_{cs}(4459)^0$ &0.84 & 0.88 & 0.88 & 0.73 & 0.73 & 0.64 & 0.73 & 0.73 & 0.62 & 0.40  \\
\noalign{\smallskip}\toprule[0.8pt]
\end{tabular}
\end{table*}

One can find from Table~\ref{ours} that the mass of the state $uudc\bar{c}$ with
$\frac{1}{2}\frac{1}{2}^-$ and pentagon structure is 4303 MeV, which is very close
to the experimental data of the state $P_c(4312)^+$. In this way, its main component
can be described as the compact state $uudc\bar{c}$ with $\frac{1}{2}\frac{1}{2}^-$
and pentagon structure in the model. No matter what its structure is, the state seems
to prefer the spin-parity assignment of $\frac{1}{2}^-$ in many theoretical frameworks
~\cite{4312moleculeqcdsum,4312moleculesbseqution,4312moleculeqdcsm,4312diquarksum,43121/2}.
Conversely, the other spin-parity assignments of $\frac{1}{2}^+$~\cite{hadrocharmonium1,
hadrocharmonium2} and $\frac{3}{2}^-$~\cite{4312-3/2} were also proposed.

The mass of the state $uudc\bar{c}$ with $\frac{1}{2}\frac{1}{2}^-$ and diquark structure
is 4344 MeV, which is highly consistent with the experimental data of the state
$P_c(4337)^+$. The states $P_c(4312)^+$ and $P_c(4337)^+$ have the same assignment of
spin-parity in the model. However, the state $P_c(4312)^+$ is pentagon structure while the
state $P_c(4337)^+$ is diquark structure. Therefore, they should be so-called QCD isomers
in the model. For the two states, Yan et al proposed three possible explanations~\cite{4337yan}:
the state $P_c(4337)^+$ is a $\chi_{c0}p$ bound state with $\frac{1}{2}^+$; the state $P_c(4337)^+$
is a $\bar{D}\Sigma_c$ molecule with $\frac{1}{2}^-$ while the state $P_c(4312)^+$ is a
$\bar{D}^*\Lambda_c$ molecule with $\frac{1}{2}^-$ or $\frac{3}{2}^-$; the states $P_c(4312)^+$
and $P_c(4337)^+$ are the coupled channel systems $\bar{D}^*\Lambda_c$ -$\bar{D}\Sigma_c$ with
$\frac{1}{2}^-$ and $\bar{D}^*\Lambda_c$ -$\bar{D}\Sigma^*_c$ with $\frac{3}{2}^-$, respectively.
Nakamura et al described the states $P_c(4312)^+$ and $P_c(4337)^+$ as interfering $\bar{D}\Sigma_c$
and $\bar{D}^*\Lambda_c$ cusps with $\frac{1}{2}^-$~\cite{thresholdcusp}.

The states $uudc\bar{c}$ with $\frac{1}{2}\frac{3}{2}^-$ and pentagon and diquark structure
have masses of $4369$ MeV and 4405 MeV in the MCFTM, respectively, both of which are in agreement 
with the experimental data of the state $P_c(4380)^+$. The model therefore approves the 
description of the state as the compact state $uudc\bar{c}$ with $\frac{1}{2}\frac{3}{2}^-$ 
and pentagon or diquark structure. The molecule structure~\cite{4380molecule}, the diquark 
structure~\cite{4380diquark}, and the diquark-triquark structure~\cite{4380diquark-triquark} 
in various theoretical frameworks also supported the spin-parity assignment of $\frac{3}{2}^-$. 
In addition, the state $uudc\bar{c}$ with $\frac{1}{2}\frac{1}{2}^-$ and octet structure 
is around 4402 MeV, which is not far away from that of the state $P_c(4380)^+$. We can not 
rule out the possibility that the main component of the state $P_c(4380)^+$ may be the state 
$uudc\bar{c}$ with $\frac{1}{2}\frac{1}{2}^-$
and octet structure.

The states $uudc\bar{c}$ with $\frac{1}{2}\frac{3}{2}^-$ and diquark and octet structures
have masses of 4405 MeV and 4473 MeV, respectively, which are not far from the experimental
data of the states $P_c(4440)^+$ and $P_c(4457)^+$. The deviations from their experimental
central data are about 35 MeV and 18 MeV, respectively. In this way, the main components of 
the states $P_c(4440)^+$ and $P_c(4457)^+$ can be described as the compact states $uudc\bar{c}$
with diquark and octet structures in the model, respectively. However, they share the same
isospin-spin-parity $\frac{1}{2}\frac{3}{2}^-$. Until now, even if ignoring their structures,
their spin-parity have been highly controversial in the various theoretical frameworks, 
such as $\frac{3}{2}^-$ and $\frac{1}{2}^-$~\cite{4312moleculeqdcsm}, $\frac{1}{2}^-$ and 
$\frac{3}{2}^-$~\cite{43121/2}, $\frac{3}{2}^+$ and $\frac{5}{2}^+$~\cite{4312-3/2},
$\frac{3}{2}^-$ and $\frac{1}{2}^+$~\cite{burns}, etc. Liu et al suggested that the discovery
of the strange pentaquark molecular state $\bar{D}^{(*)}\Xi_{c}$ may be propitious to determine
the spin of the states $P_c(4440)$ and $P_c(4457)$ in the molecular picture~\cite{spin}.

The strange state $nnsc\bar{c}$ with $0\frac{1}{2}^-$ and pentagon structure has a mass
of $4463$ MeV in the MCFTM, which is completely consistent with the experimental value of the
state $P_{cs}(4459)^0$. Hence, the model supports the interpretation of the state as the compact
state $nnsc\bar{c}$ with $0\frac{1}{2}^-$ and pentagon structure. Chiral quark model can describe
the state as $\Xi_c'\bar{D}$ molecule with $0\frac{1}{2}^-$~\cite{pcsping}. Regardless of the
$\bar{D}^*\Xi_c$ molecular and diquark pictures, QCD sum rule supports that the spin-parity
assignment of the state is $\frac{1}{2}^{-}$~\cite{pcschenhx,pcswangzg,pcssundu}. Conversely,
both of one-boson-exchange model and quasipotential Bethe-Salpeter equation favored the
interpretation of the state as the molecular picture with $\frac{3}{2}^-$~\cite{pcschenr,pcshej}.
Furthermore, Du et al favored  the actual existence of two resonances with spin $\frac{1}{2}$ 
and $\frac{3}{2}$ in the energy region of the state $P_{cs}(4459)^0$ in relation with the 
heavy-quark-spin symmetry~\cite{pcsdu}. In the hadro-charmonium model, the state $P_{cs}(4459)^0$
prefers the spin-parity assignment of $\frac{1}{2}^-$ or $\frac{3}{2}^-$~\cite{hadrocharmonium2}.

\subsection{Other $P_c$, $P_{cs}$, $P_{css}$ and $P_{csss}$ states predicted by the MCFTM}

We can describe the hidden charmed states $P_c(4312)^+$, $P_c(4337)^+$, $P_c(4380)^+$,
$P_c(4440)^+$ and $P_c(4457)^+$ as the lower spin and lower isospin members in the $P_c$
family with various color configurations. We predict other possible states $qqqc\bar{c}$ 
with high spin $S=\frac{5}{2}$ and high isospin $I=\frac{3}{2}$ in the model. One can find 
from Table~\ref{hiddencharm} that the masses of the states with $\frac{1}{2}\frac{5}{2}^-$ 
are in the range of 4516 MeV to 4616 MeV. The masses of the states with $I=\frac{3}{2}$ 
spans from 4532 MeV to 4743 MeV, which changes with their spin and color flux-tube structures.
Most of the states with $I=\frac{3}{2}$ are far away from their highest threshold
$\Sigma_c^*D^*$.

Like the $P_c$ family, the lowest state $P_{cs}(4459)^0$ indicates that there probably exist
other members in the $P_{cs}$ family. In the model, the other two $P_{cs}$ states with
$0\frac{1}{2}^-$ and diquark and octet structures have masses of around 4500 MeV, see
Table~\ref{hiddencharm}. The masses of the states with $0\frac{3}{2}^-$ and three different
structures range from 4570 MeV to 4610 MeV. The states with $1\frac{1}{2}^-$ and $1\frac{3}{2}^-$
are higher about 100 MeV than the states with $0\frac{1}{2}^-$ and $0\frac{3}{2}^-$, respectively.
Conversely, the states with $1\frac{5}{2}^-$ are lower several tens MeV than the states with
$0\frac{5}{2}^-$ because the diquark $[nn]$ with $I=1$ is in color $\bar{\mathbf{3}}$ while
the diquark $[nn]$ with $I=0$ is in color $\mathbf{6}$. In general, the interaction in color
$\bar{\mathbf{3}}$ is attractive while that in color $\mathbf{6}$ is repulsive.
The states $nnsc\bar{c}$ with $\frac{5}{2}^-$ are far away from the highest threshold
$\Xi_c^*D^*$.

The wave functions of the states states $P_{css}$ and those of the states $P_{cs}$ with $I=1$ 
have the same symmetry. Their mass difference, about 150 MeV, mainly come from the mass of 
$s$-quark. For the states $P_{css}$ with $\frac{1}{2}^-$, their masses are in the range of 
4730 MeV to 4784 MeV, which are close to the result, $4600\pm175$ MeV, predicted by the QCD 
sum rule method~\cite{pcssprediction}. Wang et al predicted double strangeness molecular 
states $\Xi_c^*\bar{D}_s^*$ with $\frac{5}{2}^-$ and $\Xi_c^{'}\bar{D}_s^*$ with
$\frac{3}{2}^-$~\cite{pcssprediction2}, which are much lower about 200 MeV than our results.
The states $nssc\bar{c}$ with $\frac{5}{2}^-$ are far away from the highest threshold
$\Omega_c^*D^*$ in the MCFTM.

The masses of the $P_{csss}$ states are higher 400 MeV than those of the states $P_{c}$ with
$I=\frac{3}{2}$ also because of the mass of $s$-quark  in the MCFTM. They are in the range 
of 4985 MeV to 5140 MeV and do not dramatically change with spin and color structures. All 
of the states $sssc\bar{c}$ are far away from the threshold $\Omega_c^*D_s^*$.

\section{summary}

The observation of the hidden charmed pentaquark states $P_c$ and $P_{cs}$ by the LHCb
Collaboration presents an extremely interesting spectrum. Their masses locate around 
the baryon-meson thresholds. However, there has not been a general consensus regarding 
their natures and structures until now. The baryon-meson molecular interpretation is
the most popular one.  

In this work, we make a systematical dynamical investigation on the hidden charm 
pentaquark states with the help of the high precision numerical method GEM in the 
multiquark color flux-tube model. The model involves the multi-body confinement 
potential based on the color flux-tube picture in the lattice QCD. Different color 
structures, pentagon, diquark and octet structure, induce the QCD isomers, which 
have the close masses in the model. Like the color-magnetic interaction, such color 
structure effect can also induce mass splitting in the spectrum and make hadron world 
more fantastic. 

The model shows a novel picture for the $P_c$ and $P_{cs}$ states. It can describe 
the states as the compact pentaquark states with different structures. The spin-parity 
of the group of $P_c(4312)^+$ and $P_c(4337)^+$ is $\frac{1}{2}^-$ while that of the 
group of $P_c(4380)^+$, $P_c(4440)^+$ and $P_c(4457)^+$ is $\frac{3}{2}^-$. Their structures
are pentagon, diquark, pentagon, diquark, and octet, respectively. The members in
each group can be analogically called QCD isomers because of their the same spin-parity
and quark content but different color structures. The singlet $P_{cs}(4459)^0$ has
pentagon structure and spin-parity of $\frac{1}{2}^-$. The structure coupling effect
in the QCD isomers should occur, which will be taken into account in the future.
Note that our model conclusion just serves as one of possible theoretical suggestions.
Proper identification of the structure and property of the states require more experimental
and theoretical scrutiny. In addition, we also predict the $P_{cs}$, $P_{c ss}$ and $P_{csss}$
families in the model. We hope that these states can be searched in experiments in the future.

The five-body confinement potential, a collective degree of freedom, binds quarks to form
the compact pentaquark states. It may shed light on our understanding of how quarks and
gluons establish hadrons in the low-energy strong interactions.

\acknowledgments

{Author thanks Prof. S.L. Zhu for helpful discussions. This research is partly supported by
the Chongqing Natural Science Foundation under Project No. cstc2019jcyj-msxmX0409 and
Fundamental Research Funds for the Central Universities under Contracts No. SWU118111.}


\begin{thebibliography}{99}
\bibitem{history} M. Gell-Mann, Phys. Lett. \textbf{8}, 214 (1964).
\bibitem{light} M. Amaryan, Arxiv: 2201.04885 [hep-ex].
\bibitem{wjju} J.J. Wu, R. Molina, E. Oset, and B.S. Zou, Phys. Rev. Lett. \textbf{105}, 232001 (2010).
\bibitem{wlwang} W.L. Wang, F. Huang, Z.Y. Zhang, and B.S. Zou, Phys. Rev. C \textbf{84}, 015203 (2011).
\bibitem{jjwu1} J.J. Wu, T.S. H. Lee, and B.S. Zou, Phys. Rev. C \textbf{85}, 044002 (2012).
\bibitem{cwxiao} C.W. Xiao, J. Nieves, and E. Oset, Phys. Rev. D \textbf{88}, 056012 (2013).
\bibitem{karliner} M. Karliner and J.L. Rosner, Phys. Rev. Lett. \textbf{115}, 122001 (2015).
\bibitem{pc43804450} R. Aaij et al. (LHCb Collaboration), Phys. Rev. Lett. \textbf{115}, 072001 (2015).
\bibitem{pc4312} R. Aaij et al. (LHCb Collaboration), Phys. Rev. Lett. \textbf{122}, 222001 (2019).
\bibitem{pcs4459} R. Aaij et al. (LHCb Collaboration), Sci.Bull. \textbf{66}, 1278 (2021).
\bibitem{pc4337} R. Aaij et al. (LHCb Collaboration), Phys. Rev. Lett. \textbf{128}, 062001 (2022).
\bibitem{moleculedu} M.L. Du, V. Baru, F.K. Guo, C. Hanhart, Ulf-G. Mei{\ss}ner, J.A. Oller,
  and Q. Wang, Phys. Rev. Lett. \textbf{124}, 072001 (2020).
\bibitem{moleculewang} B. Wang, L. Meng, and S.L. Zhu, JHEP \textbf{11}, 108 (2019).
\bibitem{4312moleculehqss} M.Z. Liu, Y.W. Pan, F.Z. Peng, M.S. Sanchez, L.S. Geng, A. Hosaka,
 and M.P. Valderrama, Phys. Rev. Lett. \textbf{122}, 242001 (2019).
\bibitem{moleculehxchen} H.X. Chen, W. Chen, X. Liu, T.G. Steele, and S.L. Zhu, Phys. Rev. Lett. \textbf{115}, 172001 (2015).
\bibitem{moleculerchen} R. Chen, X. Liu, X.Q. Li, and S.L. Zhu, Phys. Rev. Lett. \textbf{115}, 132002 (2015).
\bibitem{compactsantopinto} E. Santopinto and A. Giachino, Phys. Rev. D \textbf{96}, 014014 (2017).
\bibitem{4380cftm} C.R. Deng, J.L. Ping, H.X. Huang, and F. Wang, Phys. Rev. D \textbf{95}, 014031 (2017).
\bibitem{compactrlzhu} R.L. Zhu and C.F. Qiao, Phys. Lett. B \textbf{756}, 259 (2016).
\bibitem{compactlebed} R.F. Lebed, Phys. Lett. B \textbf{749}, 454 (2015).
\bibitem{compactali} A. Ali,I. Ahmed, M.J. Aslam, A.Y. Parkhomenkod, and A. Rehman, JHEP \textbf{10}, 256 (2019).
\bibitem{4312stancu} F. Stancu, Phys. Rev. D \textbf{104}, 054050 (2021).
\bibitem{kineticguo} F.K. Guo, U.G. Meissner, W. Wang, Z. Yang, Phys. Rev. D \textbf{92}, 071502 (2015).
\bibitem{kineticliu} X.H. Liu, Q. Wang, Q. Zhao, Phys. Lett. B \textbf{757}, 231 (2016).
\bibitem{hadrocharmonium1} M.I. Eides, V.Y. Petrov, and M.V. Polyakov, Mod. Phys. Lett. A \textbf{35}, 2050151 (2020).
\bibitem{hadrocharmonium2} J. Ferretti and E. Santopinto, arXiv: 2111.08650 [hep-ph].
\bibitem{unbound} C. Fern\'{a}ndez-Ram\'{i}rez, A. Pilloni, M. Albaladejo, A. Jackura,
 V. Mathieu, M. Mikhasenko, J.A. Silva-Castro, and A.P. Szczepaniak, Phys. Rev. Lett. \textbf{123}, 092001 (2019).
\bibitem{cusp} S.X. Nakamura, PoS CHARM2020, \textbf{029} (2021).
\bibitem{reviews} Y.R. Liu,H.X.Chen, W.Chen, X.Liu,and S.L.Zhu, Prog. Part. Nucl. Phys. \textbf{107}, 237 (2019);
 N. Brambilla, S. Eidelman, C.Hanhart, A. Nefediev, C.P. Shen, C.E. Thomas, A.Vairo, and C.Z.Yuan, Phys. Rep. \textbf{873}, 1(2020);
 F.K. Guo, X.H. Liu, and S. Sakai, Prog. Part. Nucl. Phys. \textbf{112}, 103757 (2020);
 H.X. Chen, W. Chen, X. Liu, Y.R. Liu, and S.L. Zhu, arXiv: 2204.02649 [hep-ph];
 L. Meng, B. Wang, G.J. Wang, and S.L. Zhu, arXiv: 2204.08716 [hep-ph].
\bibitem{noconcensus} Y.W. Pan, M.Z. Liu, F.Z. Peng, M.S. S\'{a}nchez, L.S. Geng, and M.P. Valderrama, Phys. Rev. D \textbf{102}, 011504 (2020)
\bibitem{lattice1} T.T. Takahashi, H. Suganuma, Y. Nemoto, and H. Matsufuru, Phys. Rev. D \textbf{65}, 114509 (2002).
\bibitem{lattice2} F. Okiharu, H. Suganuma, T.T. Takahashi, Phys. Rev. Lett. \textbf{94}, 192001 (2005);
 J.L. Ping, C.R. Deng, F. Wang, and T. Goldman, Phys. Lett. B \textbf{659}, 607 (2008).
\bibitem{richard} J.M. Richard, arXiv: 1205.4326 [hep-ph].
\bibitem{cftm} C.R. Deng, H. Chen, and J.L. Ping, Phys. Rev. D \textbf{103}, 014001 (2021);
 C.R. Deng, H. Chen, and J.L. Ping, Eur. Phys. J. A \textbf{56}, 9 (2020).
\bibitem{kappa} G.S. Bali, Phys. Rev. D \textbf{62}, 114503 (2000);
 C. Semay, Eur. Phys. J. A \textbf{22}, 353 (2004);
 N. Cardoso, M. Cardoso, and P. Bicudo, Phys. Lett. B \textbf{710}, 343 (2012).
\bibitem{collapse} R.K. Bhaduri, L.E. Cohler, and Y. Nogami, Phys. Rev. Lett. \textbf{44}, 1369 (1980).
\bibitem{vijande} J. Vijande, F. Fernandez, and A. Valcarce, J. Phys. G \textbf{31}, 481 (2005).
\bibitem{flavor-dependent} J. Weinstein and N. Isgur, Phys. Rev. D \textbf{27}, 588 (1983).
\bibitem{GEM} E. Hiyama, Y. Kino, and M. Kamimura, Prog. Part. Nucl. Phys. \textbf{51} 223 (2003).
\bibitem{diquark} C.R. Deng and S.L. Zhu, arXiv: 2204.11079 [hep-ph].
\bibitem{wave8} G. Yang, J. Ping, and J. Segovia, Phys. Rev. D \textbf{99}, 014035 (2019).
\bibitem{4312moleculeqcdsum} H.X. Chen, W. Chen and S.L. Zhu, Phys. Rev. D \textbf{100}, 051501 (2019);
 J.R. Zhang, Eur. Phys. J. C \textbf{79}, 1001 (2019);
 K. Azizi, Y. Sarac, and H. Sundu, Chin. Phys. C \textbf{45}, 053103 (2021).
\bibitem{4312moleculesbseqution} J. He, Eur. Phys. J. C \textbf{79}, 393 (2019).
\bibitem{4312moleculeqdcsm} H.X. Huang, J. He and J.L. Ping, arXiv: 1904.00221 [hep-ph].
\bibitem{Yalikun} N. Yalikun, Y.H. Lin, F.K. Guo, Y. Kamiya, and B.S. Zou, Phys. Rev. D \textbf{104}, 094039 (2021).
\bibitem{4312diquarksum} Z.G. Wang, Int. J. Mod. Phys. A \textbf{35}, 2050003 (2020).
\bibitem{43121/2} R. Chen, Z.F. Sun, X. Liu, and S.L. Zhu, Phys. Rev. D \textbf{100}, 011502 (2019);
 C.W. Xiao, J. Nieves, and E. Oset, Phys. Rev. D \textbf{100}, 014021 (2019);
 A.N. Semenova, V.V. Anisovich, and A.V. Sarantsev, Eur. Phys. J. A \textbf{56}, 142 (2020).
\bibitem{4312-3/2} A. Ali and A.Y. Parkhomenko, Phys. Lett. B \textbf{793}, 365 (2019);
\bibitem{4337yan} M.J. Yan, F.Z. Peng, M.S. S\'{a}nchez, and M.P. Valderrama, arXiv: 2108.05306v1 [hep-ph].
\bibitem{thresholdcusp} S.X. Nakamura, A. Hosaka, and Y. Yamaguchi, Phys. Rev. D \textbf{104}, L091503 (2021).
\bibitem{4380molecule} R. Chen, X. Liu, X.Q. Li and, S.L. Zhu, Phys. Rev. Lett. \textbf{115}, 132002 (2015);
  H.X. Chen, W. Chen, X. Liu, T.G. Steele, and S.L. Zhu, Phys. Rev. Lett. \textbf{115}, 172001 (2015);
  J. He, Phys. Lett. B \textbf{753}, 547 (2016);
  L. Roca, J. Nieves, and E. Oset, Phys. Rev. D \textbf{92}, 094003 (2015).
\bibitem{4380diquark} L. Maiani, A.D. Polosa, and V. Riquer, Phys. Lett. B \textbf{749}, 289 (2015);
  R.F. Lebed, Phys. Rev. D \textbf{92}, 114030 (2015);
  Z.G. Wang, Eur. Phys. J. C \textbf{76}, 70 (2016).
\bibitem{4380diquark-triquark} R.F. Lebed, Phys. Lett. B \textbf{749}, 454 (2015);
  R.L. Zhu, and C.F. Qiao, Phys. Lett. B \textbf{756}, 259 (2016).
\bibitem{burns} T.J. Burns and E.S. Swanson, Phys. Rev. D \textbf{100}, 114033 (2019).
\bibitem{spin} M.Z. Liu, Y.W. Pan, and L.S. Geng, Phys. Rev. D \textbf{103}, 034003 (2021).
\bibitem{pcsping} X.H. Hu and J.L. Ping, Eur. Phys. J. C \textbf{82}, 118 (2022).
\bibitem{pcschenhx} H.X. Chen, W. Chen, X. Liu, and X.H. Liu, Eur. Phys. J. C \textbf{81}, 409 (2021).
\bibitem{pcswangzg} Z.G. Wang, Int. J. Mod. Phys. A \textbf{36}, 2150071 (2021).
\bibitem{pcssundu} K. Azizi, Y. Sarac, and H. Sundu, Phys. Rev. D \textbf{103}, 094033 (2021).
\bibitem{pcschenr} R. Chen,	Phys. Rev. D \textbf{103}, 054007 (2021);
 R. Chen, Eur. Phys. J. C \textbf{81}, 122 (2021).
\bibitem{pcshej} J.T. Zhu, L.Q. Song, and J. He, Phys. Rev. D \textbf{103}, 074007 (2021).
\bibitem{pcsdu} M.L. Du, Z.H. Guo, and J.A. Oller, Phys. Rev. D \textbf{104}, 114034 (2021).
\bibitem{4312diquark} A.N. Semenova, V.V. Anisovich, and A.V. Sarantsev, Eur. Phys. J. A \textbf{56}, 142 (2020).
\bibitem{pcssprediction} K. Azizi, Y. Sarac, and H. Sundu, arXiv: 2112. 15543v1 [hep-ph].
\bibitem{pcssprediction2} F.L. Wang, R. Chen, and X. Liu, Phys. Rev. D \textbf{103}, 034014 (2021).
\end{thebibliography}
\end{document}